\documentclass[ aps,%
floatfix,%
final,%
notitlepage,%
oneside,%
onecolumn,%
nobibnotes,%
nofootinbib,%
superscriptaddress,%
showpacs,%
]%
{revtex4}

\usepackage{epsfig}
\usepackage{axodraw}
\newcommand{\Br}{\mathrm{Br}\,}
\newcommand{\beq}{\begin{eqnarray}}\newcommand{\eeq}{\end{eqnarray}}
\newcommand{\beqa}{\begin{eqnarray*}}\newcommand{\eeqa}{\end{eqnarray*}}

\begin{document}

\title{Inclusive decays $\chi_{b0,2}\to\psi D\bar D+X$ and the duality relation}

\author{V.V. Braguta}
\email{braguta@mail.ru}
\author{A.K. Likhoded}
\email{Likhoded@ihep.ru}
\author{A.V. Luchinsky}
\email{Alexey.Luchinsky@ihep.ru}
\affiliation{Institute for High Energy Physics, Protvino, Russia}

\begin{abstract}
In this article we consider the three-particle decays $\chi_{b0,2}\to J/\psi c\bar c\to J/\psi D\bar D+X$. We present the analytical formulae for the differential widths of these decays, the numerical values of their integrated widths and check the duality relation that connects the decays $\chi_{b0,2}\to\psi c\bar c$ with the two-particle decays $\chi_{b0,2}\to J/\psi (c\bar c)$. We also study the possibility of observing the $\chi_{b0,2}\to J/\psi DD+X$ mode at Tevatron and LHC colliders.
\end{abstract}

\pacs{%
  13.25.Gv    
  , 12.38.Bx  
  ,12.40.Nn   
  }

\maketitle

\section{Introduction}

The bound states of heavy quarkonia (for example $J/\psi$-, $\eta_c$- or $\chi_{b0,2}$-mesons) are of great interest from both theoretical and experimental points of view \cite{Brambilla:2004wf}. This interest is caused by the fact that these mesons can be simply separated at the experiment. For example, a clear signal for $J/\psi$-meson (in what follows we will denote it by $\psi$) is its leptonic decay $\psi\to e^+e^-$. On the other hand the non-relativistic nature of these mesons simplifies their theoretical description noticeably. For the calculation of probability of the processes with such mesons one can often neglect the internal motion of quarks inside mesons in the hard part of the amplitude (in what follows we will refer to this assumption as "$\delta$-approximation"). However the application of the $\delta$-approximation in two-particle processes is not valid in some cases and can lead to large errors. The cross-section of the $e^+e^-\to\psi\eta_c$ at $\sqrt{s}=10.6$ GeV obtained with the help of this approximation \cite{Braaten:2002fi}, for example, is about an order of magnitude smaller than the experimental value \cite{Abe:2002rb}. In the papers cite{Ma:2004qf,Bondar:2004sv,Braguta:2005gw} it was shown that taking into account the internal motion of quarks inside mesons one increases the theoretical predictions significantly and the agreement with the experimental results can be achieved. The physical reason of this effect is that the intermediate particles have large virtuality when the $\delta$-approximation is used, and the internal motion of quarks decreases this virtuality and hence raises the value of the cross section. The same enhancement can be also observed in excited charmonia production \cite{Braguta:2005kr}.

The method used in \cite{Ma:2004qf,Bondar:2004sv,Braguta:2005gw} can, however, lead to large uncertainties. The reason is that the quarks distribution functions play a significant role in this method and varying this functions we can change the result noticeably \cite{Ma:2004qf,Braguta:2005gw}. So, an independent way to determine these cross sections or, at least, to set some bounds on their values is desirable. This bounds can be determined with the help of duality relation that links the cross section of three-particle process $e^+e^-\to\psi c\bar c$ with the sum of two-particle cross-sections \cite{Kiselev:1994pu}:
\beq
\int\limits_{2m_c}^{M_\mathrm{th}} dm_{c\bar c}\frac{d\sigma(e^+e^-\to\psi c\bar c)}{dm_{c\bar c}}
  & \approx &
\sum\limits_\mathcal{M}\sigma(e^+e^-\to\mathcal{M}\psi).
\label{dual}
\eeq
The integration on the left hand side of this relation with respect to the invariant $c\bar c$-pair mass is held over the duality region
\beqa
2m_c<m_{c\bar c}<M_\mathrm{th}\approx 2m_D
\eeqa
and the summation on the right hand side is performed over all charmonium mesons, whose masses lie in that region (i.e. $\mathcal{M}=\eta_c,\psi,\eta_c',\chi_{c0},\dots$). It should be noticed that the cross-section $\sigma(e^+e^-\to\psi c\bar c)$ in the left hand side of the duality relation has only mild dependence on the choice of the distribution function \cite{Martynenko:2005sf} since, contrary to the two-particle reactions $e^+e^-\to\mathcal{M}\psi$, the virtualities of intermediate particles in this process are not fixed.

Besides the mentioned above reactions there are also other processes where the intermediate particles have large virtuality when the $\delta$-approximation is used. In the paper \cite{Braguta:2005gw} we considered the decays $\chi_{b0,2}\to\psi\psi$ and showed that the widths of these decays are increased when the internal motion of quarks in mesons is taken into account and depend strongly on the choice of the distribution functions. In the present article we will consider the reactions $\chi_{b0,2}\to\psi c\bar c$ that are dual to these decays and check the duality relation analogous to eq.(\ref{dual}) in this case.

As it was mentioned above, when one abandon the $\delta$-approximation, the widths of the decays $\chi_{b0,2}\to\psi\psi$ are increased. In \cite{Braguta:2005gw} we have estimated the cross-section of hadronic process $p\bar p\to\chi_{b0,2}X\to\psi\psi X$ and show that it is quite possible to use the $\chi_{b0,2}\to\psi\psi$-mode for the detection of $\chi_{bJ}$-mesons and to separate this signal from the background of the production of $\psi\psi$ pairs at Tevatron and LHC colliders. Here we will perform this analysis for the $\chi_{b0,2}\to\psi c\bar c$ decay.

The rest of our paper is organized as follows. In the next section we will describe the formalism that was used to obtain the differential $\chi_{b0,2}\to\psi c\bar c$ and present the analytical results. In the section 3 the duality relation is checked and total widths of the decays $\chi_{b0,2}\to\psi D\bar D$ are given. Section 4 is devoted to the possibility of observation of these decays at Tevatron and LHC colliders. Finally we discuss our results.

\section{\protect$\chi_{b0,2}\to\psi c\bar c$}

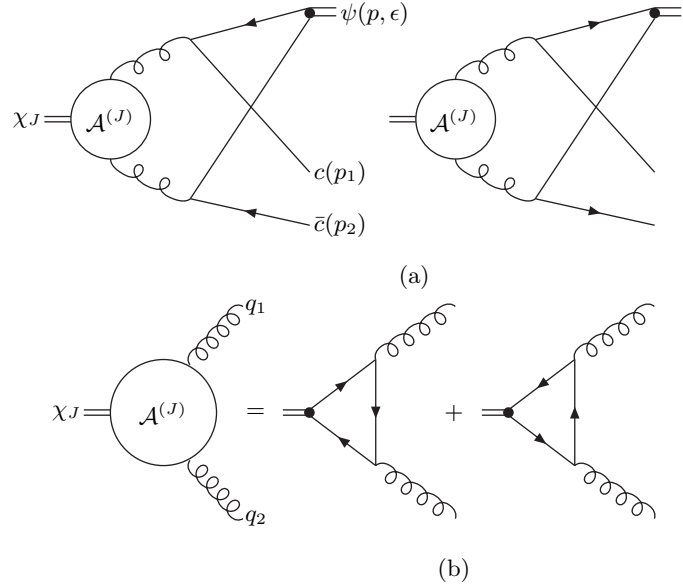
\begin{figure}

\begin{center}\begin{picture}(260,100)(0,-10)
\Line(10,51)(20,51)\Line(10,49)(20,49)\Text(9,50)[rc]{$\chi_J$}
\Oval(35,50)(15,15)(0)\Text(35,50)[cc]{$\mathcal{A}^{(J)}$}
\Gluon(35,35)(65,20){-3}{2}\Gluon(35,65)(65,80){3}{2}
\ArrowLine(110,92)(65,80)\Line(65,80)(110,30)\ArrowLine(110,10)(65,20)\Line(65,20)(110,88)
\Text(112,10)[lc]{${\bar c}(p_2)$}\Text(112,30)[lc]{$c(p_1)$}
\Vertex(110,90){2}\Line(110,92)(120,92)\Line(110,89)(120,89)
\Text(122,90)[lc]{$\psi(p,\epsilon$)}
\Line(140,51)(150,51)\Line(140,49)(150,49)
\Oval(165,50)(15,15)(0)\Text(165,50)[cc]{$\mathcal{A}^{(J)}$}
\Gluon(165,35)(195,20){-3}{2}\Gluon(165,65)(195,80){3}{2}
\ArrowLine(195,80)(240,92)\Line(195,80)(240,30)\ArrowLine(195,20)(240,10)\Line(195,20)(240,88)
\Vertex(240,90){2}\Line(240,92)(250,92)\Line(240,89)(250,89)
\Text(150,-5)[t]{(a)}
\end{picture}\end{center}

\begin{center}\begin{picture}(230,100)(0,-10)
\Line(10,51)(20,51)\Line(10,49)(20,49)\Text(9,50)[rc]{$\chi_J$}
\Oval(40,50)(20,20)(0)\Text(40,50)[cc]{$\mathcal{A}^{(J)}$}
\Gluon(50,32)(70,10){-3}{4}\Text(71,90)[l]{$q_1$}
\Gluon(50,68)(70,90){3}{4}\Text(71,10)[l]{$q_2$}
\Text(75,50)[c]{=}
\Line(85,51)(95,51)\Line(85,49)(95,49)\Vertex(95,50){2}
\ArrowLine(95,51)(120,70)\ArrowLine(120,70)(120,30)\ArrowLine(120,30)(95,49)
\Gluon(120,70)(150,90){3}{4}\Gluon(120,30)(150,10){3}{4}
\Text(150,50)[c]{+}
\Line(160,51)(170,51)\Line(160,49)(170,49)\Vertex(170,50){2}
\ArrowLine(170,49)(195,30)\ArrowLine(195,30)(195,70)\ArrowLine(195,70)(170,51)
\Gluon(195,70)(225,90){3}{4}\Gluon(195,30)(225,10){3}{4}
\Text(150,-5)[t]{(b)}
\end{picture}\end{center}
\caption{The diagrams for $\chi_{J}\to\psi c\bar c$ process and$\chi_J\to g^*g^*$ vertex}\label{figPsiCC}
\end{figure}

In what follows we will use the valence approximation, i.e. restrict ourselves to the main term in the meson fock expansion and, in particular, neglect the contribution from the color-octet states. In this approximation it is supposed that the meson consists of the color-singlet quark-antiquark pair and the probabilities of such formation are determined from the mean values of the four-fermion operators $\left<\mathcal{O}_1\right>_{\psi,\chi}$. A detailed description of the formalism used in our work can be found, for instance, in \cite{Braaten:2002fi}.

The diagrams of the process $\chi_{b0,2}\to\psi c\bar c$ in this approximation are shown on fig.\ref{figPsiCC}a. The corresponding amplitudes can be written in the form
\beqa
\mathcal{M}^{(J)} & = & \frac{8\pi\alpha_s}{q_1^2 q_2^2}\left(\lambda^a\lambda^b\right)_{ij}
  \mathcal{A}^{(J)}_{\alpha\beta}B M_\psi
  \bar u(p_1)\gamma^\alpha\hat\epsilon\left(M_\psi+\hat P\right)\gamma^\beta v_j(p_2),
\eeqa
where $p_{1,2}$ are the momenta of final quark and antiquark, $i,j$ are their color indices, $\bar u_i(p_1)$ and $v_j(p_2)$  are the corresponding wave functions, $P$ and $\epsilon$ are the momentum and polarization vector of final $\psi$-meson and $\mathcal{A}^{(J)}_{\alpha\beta}\delta^{ab}$ denotes the amplitude of the decay of initial $\chi_{bJ}$-meson into a pair of virtual gluons with the momenta $q_{1,2}$ and color indecies $a,b$ (the corresponding diagrams are shown in fig.\ref{figPsiCC}b). The widths of the $\chi_{bJ}$-mesons decays in the considered mode can be expressed through these expressions with the help of well known Dalitz relation:
\beq
\frac{d^2\Gamma(\chi_{bJ}\to\psi c\bar c)}{ds_1 ds_{12}} &=& 
  \frac{1}{256\pi^3 M_\chi^3}\overline{\left|\mathcal{M}^{(J)}\right|^2}.
  \label{Dalitz}
\eeq
Here we use the invariant masses

\beqa
s_{1,2}=M_\chi^2 x_{1,2}=(p_{1,2}+P)^2,&&\quad s_{12}=M_\chi^2 x_{12}=(p_1+p_2)^2,
\eeqa
that are linked by the relation $s_1+s_2+s_{12}=M_\chi^2+M_\psi^2+2m_c^2$. The bar in eq.(\ref{Dalitz}) stands for the average over the polarization of the initial particle (in the case of tensor meson, $J=2$) and summation over the polarization of final particles. With the help of this relations the following expression for the differential width of the decay $\chi_{b0}\to\psi c\bar c$ was obtained:
\beq
\frac{d\Gamma(\chi_{b0}\to\psi c\bar c)}{dx_1dx_{12}} & = & 
  \frac{131072\pi\alpha_s^4 C_F\xi}{9N_c^3(4x_1-\xi^2)^2(4x_2-\xi^2)^2(4+\xi^2-2x_1-2x_2)^4}
\times\nonumber\\&\times&
  \frac{\left<\mathcal{O}_1\right>_\psi \left<\mathcal{O}_1\right>_\chi}{M_\chi^4 M_\psi^3}
  \sum\limits_{n=0}^6 \xi^{2n}\mathcal{C}^{(0)}_n,
\label{dBr0}
\eeq
where $N_c=3$, the color factor $C_F=(N_c^2-1)/2N_c$, $\xi=M_\psi/M_\chi$, the mean values of the four-fermion operators $\left<\mathcal{O}_1\right>_{\psi,\chi}$ can be expressed via the values of the mesons' wave functions in the origin (see next section for details) and the coefficients $\mathcal{C}^{(0)}_n$ are defined as
\begin{eqnarray*}
\mathcal{C}^{(0)}_0 & = & -256{x_1}( 16{{x_1}}^5 + 48{{x_1}}^4( -1 + {x_{12}} )  - 96{{x_1}}^2{( -1 + {x_{12}} ) }^2 + 
 \\ & & 
   8{{x_1}}^3( 11 - 16{x_{12}} + 5{{x_{12}}}^2 )  + 
 \\ & & 
   {( 5 + {x_{12}} ) }^2( -1 + 3{x_{12}} - 11{{x_{12}}}^2 + 9{{x_{12}}}^3 )  + 
 \\ & & 
   {x_1}( 65 - 152{x_{12}} + 302{{x_{12}}}^2 + 72{{x_{12}}}^3 + {{x_{12}}}^4 )  ) ,
\\
\mathcal{C}^{(0)}_1 & = & 64( 288{{x_1}}^5 + 16{{x_1}}^4( -11 + 51{x_{12}} )  + 
 \\ & & 
   32{{x_1}}^3( -1 - 20{x_{12}} + 21{{x_{12}}}^2 )  + 
 \\ & & 
   {( 5 + {x_{12}} ) }^2( 3 + 3{x_{12}} - 23{{x_{12}}}^2 + {{x_{12}}}^3 )  + 
 \\ & & 
   16{{x_1}}^2( 30 + 201{x_{12}} + 62{{x_{12}}}^2 + 19{{x_{12}}}^3 )  + 
 \\ & & 
   2{x_1}( -205 - 1296{x_{12}} + 1394{{x_{12}}}^2 + 864{{x_{12}}}^3 + 107{{x_{12}}}^4 )  ) ,
\\
\mathcal{C}^{(0)}_2 & = & -16( -1515 + 2096{{x_1}}^4 - 7688{x_{12}} - 3562{{x_{12}}}^2 - 392{{x_{12}}}^3 + 5{{x_{12}}}^4 + 
 \\ & & 
   32{{x_1}}^3( 73 + 167{x_{12}} )  + 16{{x_1}}^2( 151 + 304{x_{12}} + 261{{x_{12}}}^2 )  + 
 \\ & & 
   128{x_1}( -16 + 141{x_{12}} + 92{{x_{12}}}^2 + 17{{x_{12}}}^3 )  ) ,
\\
\mathcal{C}^{(0)}_3 & = & 64( -1180 + 492{{x_1}}^3 - 1125{x_{12}} - 316{{x_{12}}}^2 - 11{{x_{12}}}^3 + 
 \\ & & 
   2{{x_1}}^2( 493 + 515{x_{12}} )  + 2{x_1}( 833 + 980{x_{12}} + 335{{x_{12}}}^2 )  ) ,
\\
\mathcal{C}^{(0)}_4 & = & -8( -3961 + 1742{{x_1}}^2 - 3152{x_{12}} - 319{{x_{12}}}^2 + {x_1}( 4258 + 2654{x_{12}} )  ) ,
\\
\mathcal{C}^{(0)}_5 & = & 4( -2479 + 258{x_1} - 825{x_{12}} ) ,
\\
\mathcal{C}^{(0)}_6 & = & 1515.
\end{eqnarray*}
For the decay $\chi_{b2}\to\psi c\bar c$ we have
\beq
\frac{d\Gamma(\chi_{b2}\to\psi c\bar c)}{dx_1dx_{12}} & = & 
  \frac{262144\pi\alpha_s^4 C_F\xi}{45N_c^3(4x_1-\xi^2)^2(4x_2-\xi^2)^2(4+\xi^2-2x_1-2x_2)^4}
\times\nonumber\\&\times&
  \frac{\left<\mathcal{O}_1\right>_\psi \left<\mathcal{O}_1\right>_\chi}{M_\chi^4 M_\psi^3}
  \sum\limits_{n=0}^6 \xi^{2n}\mathcal{C}^{(2)}_n,
  \label{dBr2}
\eeq
\begin{eqnarray*}
\mathcal{C}^{(2)}_0 & = & -256{x_1}( -1 + {x_1} + {x_{12}} ) ( 49 + 16{{x_1}}^4 + 32{{x_1}}^3( -1 + {x_{12}} )  + 
 \\ & & 
   8{{x_1}}^2( 1 + ( -20 + {x_{12}} ) {x_{12}} )  - 
 \\ & & 
   8{x_1}( -1 + {x_{12}} ) ( 1 + {x_{12}}( 16 + {x_{12}} )  )  + 
 \\ & & 
   {x_{12}}( -100 + {x_{12}}( 158 + {x_{12}}( -20 + 9{x_{12}} )  )  )  ) ,
\\
\mathcal{C}^{(2)}_1 & = & 64( 147 + 288{{x_1}}^5 + 429{x_{12}} - 266{{x_{12}}}^2 + 98{{x_{12}}}^3 - 25{{x_{12}}}^4 + {{x_{12}}}^5 + 
 \\ & & 
   16{{x_1}}^4( -47 + 51{x_{12}} )  + 32{{x_1}}^3( 17 - 92{x_{12}} + 21{{x_{12}}}^2 )  + 
 \\ & & 
   16{{x_1}}^2( -30 + 261{x_{12}} - 154{{x_{12}}}^2 + 19{{x_{12}}}^3 )  + 
 \\ & & 
   {x_1}( 694 - 3144{x_{12}} + 3460{{x_{12}}}^2 - 648{{x_{12}}}^3 + 214{{x_{12}}}^4 )  ) ,
\\
\mathcal{C}^{(2)}_2 & = & -16( 1149 + 2096{{x_1}}^4 - 4148{x_{12}} + 1622{{x_{12}}}^2 - 452{{x_{12}}}^3 + 5{{x_{12}}}^4 + 
 \\ & & 
   32{{x_1}}^3( -143 + 167{x_{12}} )  + 16{{x_1}}^2( 493 - 776{x_{12}} + 261{{x_{12}}}^2 )  + 
 \\ & & 
   32{x_1}( -253 + 645{x_{12}} - 172{{x_{12}}}^2 + 68{{x_{12}}}^3 )  ) ,
\\
\mathcal{C}^{(2)}_3 & = & 64( -772 + 492{{x_1}}^3 + 567{x_{12}} - 184{{x_{12}}}^2 - 11{{x_{12}}}^3 + 
 \\ & & 
   2{{x_1}}^2( -383 + 515{x_{12}} )  + 2{x_1}( 1115 - 544{x_{12}} + 335{{x_{12}}}^2 )  ) ,
\\
\mathcal{C}^{(2)}_4 & = & -8( 1937 + 1742{{x_1}}^2 - 836{x_{12}} - 319{{x_{12}}}^2 + 2{x_1}( -607 + 1327{x_{12}} )  ) ,
\\
\mathcal{C}^{(2)}_5 & = & 4( -19 + 258{x_1} - 825{x_{12}} ) ,
\\
\mathcal{C}^{(2)}_6 & = & 1515.
\end{eqnarray*}

\section{Numerical results and duality relation}

Using expressions presented above for the $\chi_{b0,2}\to\psi c\bar c$ branching fractions we can obtain their numerical values. In our calculations the following values for $c$-quark, $\chi$- and $\psi$-mesons were used:
\beqa
m_c    & = & 1.5\,\mathrm{GeV},\\
M_\psi & = & 3.097\,\mathrm{GeV},\\
M_\chi & = & (M_{\chi_{b0}}+M_{\chi_{b2}})/2= 9.886\,\mathrm{GeV},
\eeqa
and the strong coupling constant was set to be $\alpha_s=0.2$. The mean value of the four-fermion operator $\left<\mathcal{O}_1\right>_\psi$ in eqs.(\ref{dBr0}), (\ref{dBr2}) is related to the wave function of the $\psi$-meson at origin or the leptonic width of this meson:
\beqa
\left<\mathcal{O}_1\right>_\psi & = & 2N_c |\psi(0)|^2 = \frac{3}{2\pi e_c^2\alpha^2}\Gamma(\psi\to e^+e^-)m_c^2
  = 0.261\,\mathrm{GeV}^3.
\eeqa
Since we do not consider loop corrections we can use only leading order results in these expressions. The matrix element $\left<\mathcal{O}_1\right>_\chi$  can be expressed through the value of the derivative of $\chi_b$-meson wave function at the origin:
\beqa
\left<\mathcal{O}_1\right>_\chi &=& \frac{3N_c}{2\pi}\left|R'(0)\right|^2.
\eeqa
In this work we use $|R'(0)|^2=1.34\,\mathrm{GeV}^5$ \cite{Olsson:1984im} and total widths of $\chi_{b0,2}$-mesons in color-singlet approximation are equal to
\beqa
\Gamma(\chi_{b0}\to gg) & = & 96\alpha_s^2\frac{|R'(0)|^2}{M_\chi^4} \approx 0.544\,\mbox{MeV},\\
\Gamma(\chi_{b0}\to gg) & = & \frac{128\alpha_s^2}{5}\frac{|R'(0)|^2}{M_\chi^4} \approx 0.154\,\mbox{MeV}.
\eeqa
In the works \cite{Bodwin:1992ye,Brambilla:2001xy} it was shown that for P-wave mesons (for example $\chi_{bJ}$) the contribution of color-octet states is not suppressed with respect to singlet ones. Our estimates, however, show that this effect does not lead to noticeable change of $\chi_{b0,2}$ total widths (actually, $\chi_{b0}$ widths is increased by approximately 5\%, $\chi_{b2}$ width by 20\%) and we will neglect these corrections.

\begin{figure}   
\includegraphics{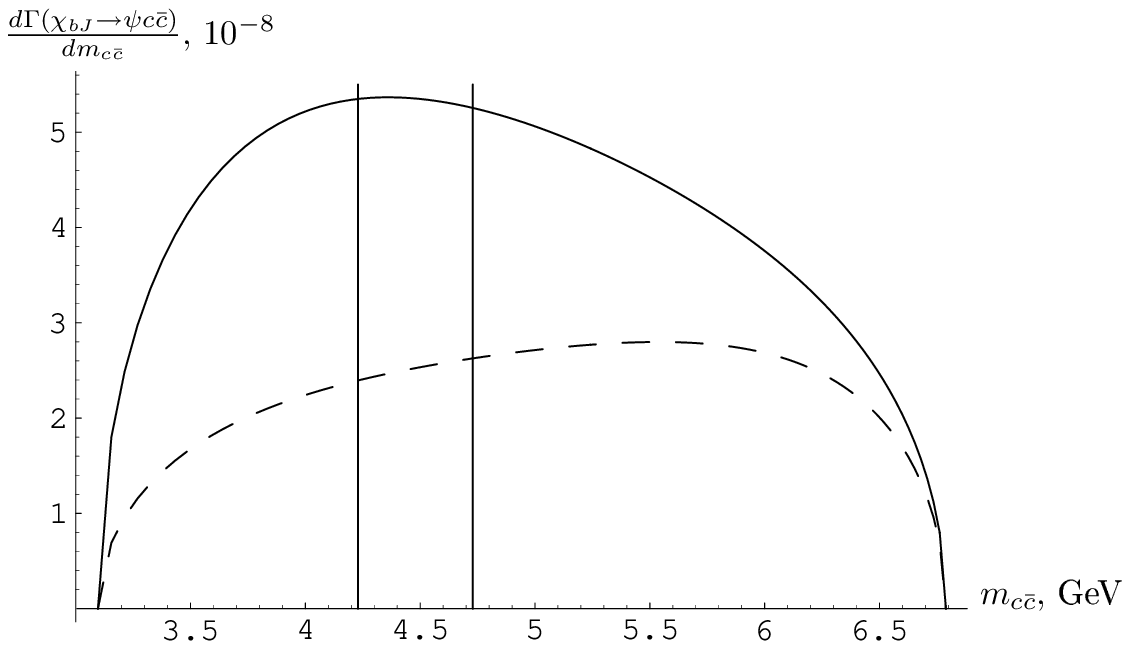}
\caption{The distribution of the $\chi_{b0}$ (dashed line) and $\chi_{b2}$ (solid line) branching fractions versus $m_{c\bar c}$}
\label{figdGdMcc}
\end{figure}

In the fig.\ref{figdGdMcc} the distributions of $\chi_{b0}\to\psi c\bar c$ (dashed line) and $\chi_{b0}\to\psi c\bar c$ (solid line) versus the invariant mass of quark-antiquark pair $m_{c\bar c}$ are shown. For the large values of $m_{c\bar c}$ the hadronization of this pair into $D\bar D$ is possible. For small values of $m_{c\bar c}$ the production of $D\bar D$ is forbidden and the quark-antiquark pair has to form a bound state, i.e. charmonium. This fact makes it possible to write a duality relation that is analogous to the one presented in \cite{Kiselev:1994pu}:
\beq
\int\limits_{2m_c}^{M_\mathrm{th}}dm_{c\bar c}\frac{dBr(\chi_{bJ}\to\psi c\bar c)}{dm_{c\bar c}} & = &
  \sum\limits_\mathcal{M} \Br(\chi_{bJ}\to\psi\mathcal{M}).
\label{dual2}
\eeq
In the left hand side of this relation the integration is held over the duality interval
\beqa
2m_c &<& m_{c\bar c} < M_\mathrm{th} = 2m_D+\delta{M},
\eeqa
and the summation on the right hand side over the $(c\bar c)$ mesons, whose masses belong to this interval. Because of charge parity conservations this should be vector mesons and excited mesons do not change the sum in the r.h.s. of eq.(\ref{dual2}) significantly, so we will use only $\mathcal{M}=\psi, \psi(2S)$ and $\psi(3770)$. We will also suppose that the excited states differ form $\psi$-meson only by the mass and the value of mesonic constant that can be determined from the width of the leptonic decay. That is why the following relation is valid:
\beqa
\Br(\chi_{bJ}\to\psi\mathcal{M}) & = & 
  \frac{1}{2}\frac{\Gamma(\mathcal{M}\to e^+e^-)}{\Gamma(\psi\to e^+e^-)}
  \frac{M_\psi}{M_\mathcal{M}} \Br(\chi_{bJ}\to\psi\psi).
\eeqa
According to \cite{Kiselev:1994pu} the reasonable values of the parameter $\delta M$ belong to the interval
\beqa
0.5\,\mbox{GeV} & < & \delta M < 1\,\mbox{GeV},
\eeqa
that is shown on fig.\ref{figdGdMcc} by vertical lines. This restriction sets the bounds on the $\chi_{b0,2}\to\psi\psi$ branching fractions:
\beqa
2.36\cdot 10^{-5} < \Br(\chi_{b0}\to\psi\psi) < 4.13\cdot 10^{-5},\\
2.12\cdot 10^{-4} < \Br(\chi_{b2}\to\psi\psi) < 3.53\cdot 10^{-4}.
\eeqa
As in was mentioned above, the theoretical predictions for this branching fractions depend strongly on the choice of meson distribution functions and the presented above restrictions exclude some variants. In particular, from all distribution considered in \cite{Braguta:2005gw} only the distribution $\phi(x)\sim x^3(1-x)^3$ \cite{Kartvelishvili:1977pi} gives the branching fractions that are in these intervals
\beqa
\Br(\chi_{b0}\to\psi\psi) & = & 2.88\cdot 10^{-5},\qquad
\Br(\chi_{b2}\to\psi\psi)   =   2.5\cdot 10^{-4}
\eeqa
Using these numbers we obtain form eq.(\ref{dual2}) the values
\beqa
\delta M = 0.75\,\mbox{GeV}
\eeqa
for the scalar meson $\chi_{b0}$ and
\beqa
\delta M = 0.74\,\mbox{GeV}
\eeqa
for $\chi_{b2}$.

At $m_{c\bar c}>M_\mathrm{th}$ the hadronization of the quark-antiquark pair into $D\bar D$-pair and light mesons is possible. We will assume that the probability of this transition is 100\% and that in this region the reactions $\chi_{bJ}\to\psi c\bar c$ and $\chi_{bJ}\to\psi D\bar D+X$ are identical. The branching fractions of these decays are equal to
\beqa
\Br(\chi_{b0}\to\psi D\bar D+X) & = & 1.19\cdot 10^{-4},\\
\Br(\chi_{b2}\to\psi D\bar D+X) & = & 7.06\cdot 10^{-4}.
\eeqa

\section{$p\bar p\to\psi D\bar D+X$}

In paper \cite{Braguta:2005gw} we considered the possibility of observation $\chi_b$-mesons in the mode $\chi_b\to\psi\psi$ at Tevatron and LHC and separation this signal from the background of non-resonance $\psi\psi$ production. In this section we will present the analogous analysis for $\chi_{bJ}\to\psi D\bar D+X$ reaction. 

The cross section of the reaction $p\bar p\to\psi D\bar D+X$ can be expressed through the cross section of the partonic subprocess
\beq
\sigma & = & \sum\limits_{a,b}\int\limits_0^1 dx_a dx_b f_{a/p}(x_a) f_{b/\bar p}(x_b)\hat\sigma(ab\to\psi c\bar c),
\label{sigma}
\eeq
where sum is over partons $a$ and $b$, $x_{a,b}$ are the longitudinal momentum fractions of these processes and $f_{a/p}$ and $f_{b/\bar p}$ are their structure functions in the initial baryons. In our calculations we used for this functions the results of the work \cite{Alekhin:2002fv}. If we are interested in central region production where $x_a=x_b=M_\chi/\sqrt{s}\ll 1$ ($\sqrt{s}$ is the energy of initial particles in the c.m.s. frame, $\sqrt{s}=2$ TeV for Tevatron and $\sqrt{s}=14$ TeV for LHC), than the main contribution to the eq.(\ref{sigma}) gives the gluonic pair (i.e. $a=b=g$) and it can be written in the form
\beq
\frac{d\sigma(p\bar p\to \psi D\bar D+X)}{d\hat s} & = & \frac{\hat\sigma_{gg}}{\hat s}L.
\label{dsigma}
\eeq
Here $\hat s=M_\chi^2 x_a x_b=M_{gg}^2$ is the squared energy of the gluonic pair in c.m.s, $\hat\sigma_{gg}=\hat\sigma(gg\to\psi D\bar D+X)$, and the dimensionless factor
\beq
L(\hat s) & = & 2\int\limits_0^{1-\hat s/s}\frac{x_a x_b}{x_a+x_b} f_{g/p}(x_a) f_{g/\bar p}(x_b) dx
\label{L}
\eeq
describes the hadronic luminosity.

\begin{figure}   
\includegraphics{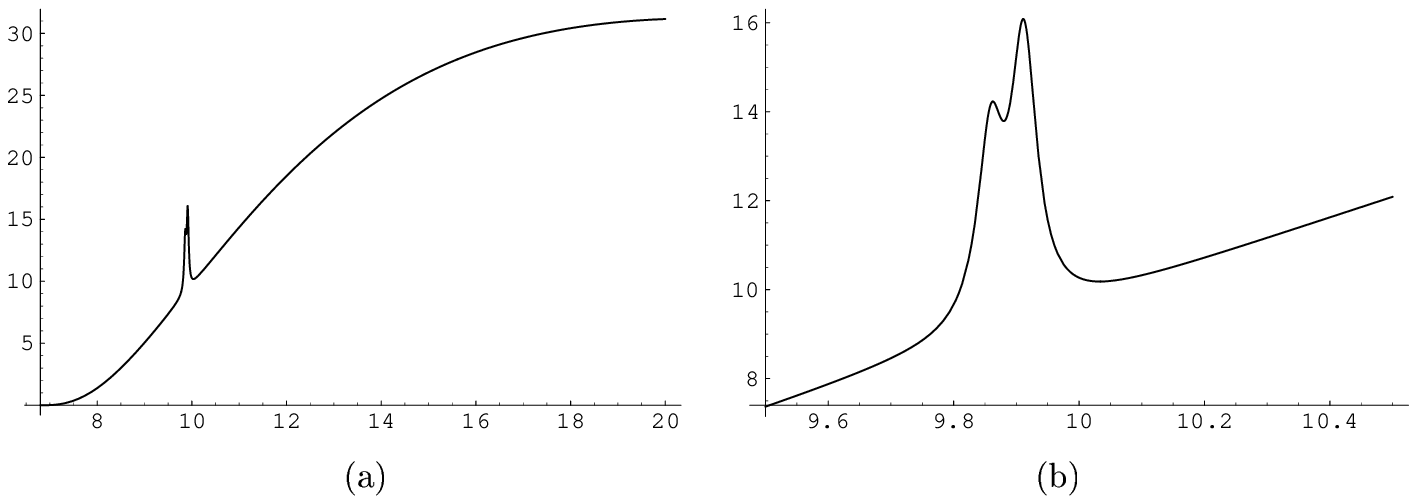}
\caption{The full partonic cross section (in pb) versus the gluon pair invariant mass $M_{gg}$ (in GeV). The value of the instrumental error is $\Delta=50$ MeV. In subfigure (b) we show the region $M_{gg}\approx M_\chi$ in detail.}
\label{figSigma}
\end{figure}

The cross section of the partonic subprocess is the sum of resonance cross sections $\hat\sigma_J=\hat\sigma(gg\to\chi_{bJ}\to\psi D\bar D+X)$ and non-resonance background. 
The latter process was studied in detail in \cite{Berezhnoi:1998aa}, where a rigorous description of $gg\to\psi c\bar c$ subprocess is presented. Since we are interested only in the $\hat s$ dependence of the cross section of this process, it is convenient to use the parameterization
\beqa
\hat\sigma_{\mathrm{nr}} & = & 
  518 \left(1-\frac{4m_c}{\sqrt{\hat s}}\right)^{3.0}\left(\frac{4m_c}{\sqrt{\hat s}}\right)^{1.45}\,\mbox{pb}
\eeqa
that can be found in that paper. The resonance cross sections can be obtained from the presented above branching fractions with the help of Breit-Wigner formula:
\beqa
\hat\sigma_J & = & \frac{(2J+1)}{16}\frac{\pi}{M_{gg}^2 }
  \frac{\Gamma(\chi_{bJ}\to gg)\Gamma(\chi_{bJ}\to\psi D\bar D+X)}{(M_{gg}-M_{\chi_{bJ}})^2+\Gamma_{\chi_{bJ}}^2/4}.
\eeqa
The widths of the resonances $\Gamma_{\chi_{bJ}}$ are, however, small in comparison with the detector instrumental error $\Delta$. We will take this error into account by the means of a simple substitution
\beqa
\hat\sigma_J^\Delta & = & \frac{(2J+1)}{16}\frac{\Delta}{\Gamma_{\chi_{bJ}}}\frac{\pi}{M_{gg}^2 }
  \frac{\Gamma(\chi_{bJ}\to gg)\Gamma(\chi_{bJ}\to\psi D\bar D+X)}{(M_{gg}-M_{\chi_{bJ}})^2+\Delta^2/4}.
\eeqa
One can easily see this substitution does not change the value of the integrated cross section (i.e. $\int dM_{gg}\hat\sigma_J=\int dM_{gg}\hat \sigma_J^\Delta$). In fig.\ref{figSigma} the dependence of the total partonic cross section on the invariant mass of the gluon pair is shown (the value of the instrumental error $\Delta=50$ MeV was used).

Let us now return to the cross section of the $\chi_{b0,2}$-meson production. Because this resonances are narrow we can neglect the dependence if the partonic luminosity $L$ on $\hat s$ on the width of the resonance and the integral in (\ref{dsigma}) can be written in the simple form
\beqa
\sum\limits_{J=0,2}\sigma(p\bar p\to\chi_{bJ}+X) & = & 
  \frac{\pi^2}{4M_\chi^3}L\sum\limits_J (2J+1)\Gamma(\chi_{bJ}\to gg),
\eeqa
where the value of the partonic luminosity $L$ is defined according to formula (\ref{L}). For the distributions of gluons in the initial (anti)protons we used the parameterization presented in the work \cite{Alekhin:2002fv} with the value of virtuality $Q^2=M_\chi^2$. Our calculations give $L(M_\chi)=450$ at Tevatron energies ($\sqrt{s}=2$ TeV) and $L(M_\chi)=2.9\cdot10^3$ for LHC ($\sqrt{s}=14$ TeV). In table I we present the cross sections of $\chi_{b0,2}$ productions and reactions $p\bar p\to\chi_{bJ}X\to\psi D\bar D+X$ at these energies.

\begin{table}
$$\begin{array}{|c|c||c|c|}
\hline
J & \sqrt{s},\,\,\mbox{TeV} & \sigma(pp\to\chi_{bJ}+X),\,\mu\mbox{b}& \sigma(pp\to\chi_{bJ}+X\to\psi D\bar D+X),\,\mbox{nb}\\
\hline
0 & 2  & 0.25 & 0.03  \\
  & 14 & 1.5  & 0.18 \\
\hline 
2 & 2  & 0.32 & 0.26 \\
  & 14 & 2    & 1.41\\
\hline
\end{array}$$
\label{tab}\caption{The cross sections $\sigma(p\bar p\to\chi_{BJ}+X)$ and $\sigma(p\bar p\to\chi_{BJ}+X\to \psi D\bar D+X)$ at Tevatron and LHC}
\end{table}

\section{Conclusion}

In the recent papers \cite{Ma:2004qf,Bondar:2004sv,Braguta:2005gw} it was shown that the internal motion of quarks in mesons increase the probabilities of two-particle reactions significantly and, in particular, an agreement between theoretical predictions and experimental values of the $e^+e^-\to\psi\eta_c$ cross section and $\chi_{c0}\to\phi\phi$ branching fraction can be reached. A poor knowledge of the distribution functions of quarks inside mesons can lead, however, to large uncertainties and an independent way to determine these probabilities or set some bounds on their values is desirable. For example, one can use the duality relation, that links the probabilities of three-particle reaction and two-particle processes that are dual to it. In our paper we have considered the three-particle process $\chi_{b0,2}\to\psi c\bar c$ and found the restrictions on the $\chi_{b0,2}\to\psi\psi$ branching fractions:
\beqa
2.36\cdot 10^{-5} < \Br(\chi_{b0}\to\psi\psi) < 4.13\cdot 10^{-5},\\
2.12\cdot 10^{-4} < \Br(\chi_{b2}\to\psi\psi) < 3.53\cdot 10^{-4}.
\eeqa
These restrictions, in particular, show, that from all distributions considered in \cite{Braguta:2005gw} only $\phi(x)\sim x^3(1-x)^3$ is allowed.

In the kinematically allowed region the decays $\chi_{b0,2}\to\psi c\bar c$ correspond to the decays $\chi_{b0,2}\to\psi D\bar D+X$. Our estimates for their branching fractions are
\beqa
\Br(\chi_{b0}\to\psi D\bar D+X) & = & 1.19\cdot 10^{-4},\\
\Br(\chi_{b2}\to\psi D\bar D+X) & = & 7.06\cdot 10^{-4}.
\eeqa

We have also considered the production of $\chi_{b0,2}$ mesons at Tevatron and LHC colliders and shown that it is possible to use the $\chi_{b0,2}\to\psi D\bar D+X$ decays to detect these mesons and to separate this reaction from the background.

\begin{acknowledgments}
This work was partially
supported by Russian Foundation of Basic Research under grant 04-02-17530, Russian Education
Ministry grant E02-31-96, CRDF grant MO-011-0, Scientific School grant SS-1303.2003.2. One of
the authors (V.B.) was also supported by Dynasty foundation.

\end{acknowledgments}


\begin{thebibliography}{15}
\expandafter\ifx\csname natexlab\endcsname\relax\def\natexlab#1{#1}\fi
\expandafter\ifx\csname bibnamefont\endcsname\relax
  \def\bibnamefont#1{#1}\fi
\expandafter\ifx\csname bibfnamefont\endcsname\relax
  \def\bibfnamefont#1{#1}\fi
\expandafter\ifx\csname citenamefont\endcsname\relax
  \def\citenamefont#1{#1}\fi
\expandafter\ifx\csname url\endcsname\relax
  \def\url#1{\texttt{#1}}\fi
\expandafter\ifx\csname urlprefix\endcsname\relax\def\urlprefix{URL }\fi
\providecommand{\bibinfo}[2]{#2}
\providecommand{\eprint}[2][]{\url{#2}}

\bibitem[{\citenamefont{Brambilla et~al.}(2004)}]{Brambilla:2004wf}
\bibinfo{author}{\bibfnamefont{N.}~\bibnamefont{Brambilla}}
  \bibnamefont{et~al.} (\bibinfo{year}{2004}), \eprint{hep-ph/0412158}.

\bibitem[{\citenamefont{Braaten and Lee}(2003)}]{Braaten:2002fi}
\bibinfo{author}{\bibfnamefont{E.}~\bibnamefont{Braaten}} \bibnamefont{and}
  \bibinfo{author}{\bibfnamefont{J.}~\bibnamefont{Lee}},
  \bibinfo{journal}{Phys. Rev.} \textbf{\bibinfo{volume}{D67}},
  \bibinfo{pages}{054007} (\bibinfo{year}{2003}),
  \eprint{hep-ph/0211085}.

\bibitem[{\citenamefont{Abe et~al.}(2002)}]{Abe:2002rb}
\bibinfo{author}{\bibfnamefont{K.}~\bibnamefont{Abe}} \bibnamefont{et~al.}
  (\bibinfo{collaboration}{Belle}), \bibinfo{journal}{Phys. Rev. Lett.}
  \textbf{\bibinfo{volume}{89}}, \bibinfo{pages}{142001}
  (\bibinfo{year}{2002}), \eprint{hep-ex/0205104}.

\bibitem[{\citenamefont{Braguta
  et~al.}(2005{\natexlab{a}})\citenamefont{Braguta, Likhoded, and
  Luchinsky}}]{Braguta:2005kr}
\bibinfo{author}{\bibfnamefont{V.~V.} \bibnamefont{Braguta}},
  \bibinfo{author}{\bibfnamefont{A.~K.} \bibnamefont{Likhoded}},
  \bibnamefont{and} \bibinfo{author}{\bibfnamefont{A.~V.}
  \bibnamefont{Luchinsky}}, \bibinfo{journal}{Phys. Rev.}
  \textbf{\bibinfo{volume}{D72}}, \bibinfo{pages}{074019}
  (\bibinfo{year}{2005}{\natexlab{a}}), \eprint{hep-ph/0507275}.

\bibitem[{\citenamefont{Ma and Si}(2004)}]{Ma:2004qf}
\bibinfo{author}{\bibfnamefont{J.~P.} \bibnamefont{Ma}} \bibnamefont{and}
  \bibinfo{author}{\bibfnamefont{Z.~G.} \bibnamefont{Si}},
  \bibinfo{journal}{Phys. Rev.} \textbf{\bibinfo{volume}{D70}},
  \bibinfo{pages}{074007} (\bibinfo{year}{2004}), \eprint{hep-ph/0405111}.

\bibitem[{\citenamefont{Bondar and Chernyak}(2005)}]{Bondar:2004sv}
\bibinfo{author}{\bibfnamefont{A.~E.} \bibnamefont{Bondar}} \bibnamefont{and}
  \bibinfo{author}{\bibfnamefont{V.~L.} \bibnamefont{Chernyak}},
  \bibinfo{journal}{Phys. Lett.} \textbf{\bibinfo{volume}{B612}},
  \bibinfo{pages}{215} (\bibinfo{year}{2005}), \eprint{hep-ph/0412335}.

\bibitem[{\citenamefont{Braguta
  et~al.}(2005{\natexlab{b}})\citenamefont{Braguta, Likhoded, and
  Luchinsky}}]{Braguta:2005gw}
\bibinfo{author}{\bibfnamefont{V.~V.} \bibnamefont{Braguta}},
  \bibinfo{author}{\bibfnamefont{A.~K.} \bibnamefont{Likhoded}},
  \bibnamefont{and} \bibinfo{author}{\bibfnamefont{A.~V.}
  \bibnamefont{Luchinsky}}, \bibinfo{journal}{Phys. Rev.}
  \textbf{\bibinfo{volume}{D72}}, \bibinfo{pages}{094018}
  (\bibinfo{year}{2005}{\natexlab{b}}), \eprint{hep-ph/0506009}.

\bibitem[{\citenamefont{Kiselev et~al.}(1994)\citenamefont{Kiselev, Likhoded,
  and Shevlyagin}}]{Kiselev:1994pu}
\bibinfo{author}{\bibfnamefont{V.~V.} \bibnamefont{Kiselev}},
  \bibinfo{author}{\bibfnamefont{A.~K.} \bibnamefont{Likhoded}},
  \bibnamefont{and} \bibinfo{author}{\bibfnamefont{M.~V.}
  \bibnamefont{Shevlyagin}}, \bibinfo{journal}{Phys. Lett.}
  \textbf{\bibinfo{volume}{B332}}, \bibinfo{pages}{411} (\bibinfo{year}{1994}),
  \eprint{hep-ph/9408407}.

\bibitem[{\citenamefont{Martynenko}(2005)}]{Martynenko:2005sf}
\bibinfo{author}{\bibfnamefont{A.~P.} \bibnamefont{Martynenko}}
  (\bibinfo{year}{2005}),
  \eprint{hep-ph/0506324}.

\bibitem[{\citenamefont{Olsson et~al.}(1985)\citenamefont{Olsson, Martin, and
  Peacock}}]{Olsson:1984im}
\bibinfo{author}{\bibfnamefont{M.~G.} \bibnamefont{Olsson}},
  \bibinfo{author}{\bibfnamefont{A.~D.} \bibnamefont{Martin}},
  \bibnamefont{and} \bibinfo{author}{\bibfnamefont{A.~W.}
  \bibnamefont{Peacock}}, \bibinfo{journal}{Phys. Rev.}
  \textbf{\bibinfo{volume}{D31}},
  \bibinfo{pages}{81}
  (\bibinfo{year}{1985}).

\bibitem[{\citenamefont{Bodwin et~al.}(1992)\citenamefont{Bodwin, Braaten, and
  Lepage}}]{Bodwin:1992ye}
\bibinfo{author}{\bibfnamefont{G.~T.} \bibnamefont{Bodwin}},
  \bibinfo{author}{\bibfnamefont{E.}~\bibnamefont{Braaten}}, \bibnamefont{and}
  \bibinfo{author}{\bibfnamefont{G.~P.} \bibnamefont{Lepage}},
  \bibinfo{journal}{Phys. Rev.} \textbf{\bibinfo{volume}{D46}},
  \bibinfo{pages}{R1914}
  (\bibinfo{year}{1992}), \eprint{hep-lat/9205006}.

\bibitem[{\citenamefont{Brambilla et~al.}(2002)\citenamefont{Brambilla, Eiras,
  Pineda, Soto, and Vairo}}]{Brambilla:2001xy}
\bibinfo{author}{\bibfnamefont{N.}~\bibnamefont{Brambilla}},
  \bibinfo{author}{\bibfnamefont{D.}~\bibnamefont{Eiras}},
  \bibinfo{author}{\bibfnamefont{A.}~\bibnamefont{Pineda}},
  \bibinfo{author}{\bibfnamefont{J.}~\bibnamefont{Soto}}, \bibnamefont{and}
  \bibinfo{author}{\bibfnamefont{A.}~\bibnamefont{Vairo}},
  \bibinfo{journal}{Phys. Rev. Lett.} \textbf{\bibinfo{volume}{88}},
  \bibinfo{pages}{012003}
  (\bibinfo{year}{2002}),
  \eprint{hep-ph/0109130}.

\bibitem[{\citenamefont{Kartvelishvili
  et~al.}(1978)\citenamefont{Kartvelishvili, Likhoded, and
  Petrov}}]{Kartvelishvili:1977pi}
\bibinfo{author}{\bibfnamefont{V.~G.} \bibnamefont{Kartvelishvili}},
  \bibinfo{author}{\bibfnamefont{A.~K.} \bibnamefont{Likhoded}},
  \bibnamefont{and} \bibinfo{author}{\bibfnamefont{V.~A.}
  \bibnamefont{Petrov}}, \bibinfo{journal}{Phys. Lett.}
  \textbf{\bibinfo{volume}{B78}}, \bibinfo{pages}{615} (\bibinfo{year}{1978}).

\bibitem[{\citenamefont{Alekhin}(2003)}]{Alekhin:2002fv}
\bibinfo{author}{\bibfnamefont{S.}~\bibnamefont{Alekhin}},
  \bibinfo{journal}{Phys. Rev.} \textbf{\bibinfo{volume}{D68}},
  \bibinfo{pages}{014002} (\bibinfo{year}{2003}),
  \eprint{hep-ph/0211096}.

\bibitem[{\citenamefont{Berezhnoi et~al.}(1998)\citenamefont{Berezhnoi,
  Kiselev, Likhoded, and Onishchenko}}]{Berezhnoi:1998aa}
\bibinfo{author}{\bibfnamefont{A.~V.} \bibnamefont{Berezhnoi}},
  \bibinfo{author}{\bibfnamefont{V.~V.} \bibnamefont{Kiselev}},
  \bibinfo{author}{\bibfnamefont{A.~K.} \bibnamefont{Likhoded}},
  \bibnamefont{and} \bibinfo{author}{\bibfnamefont{A.~I.}
  \bibnamefont{Onishchenko}}, \bibinfo{journal}{Phys. Rev.}
  \textbf{\bibinfo{volume}{D57}},
  \bibinfo{pages}{4385}
  (\bibinfo{year}{1998}), \eprint{hep-ph/9710339}.

\end{thebibliography}
\end{document}